\documentclass[12pt]{article}
\usepackage{a4wide}
\usepackage{amssymb}
\usepackage{graphicx}
\begin{document}
{\renewcommand{\thefootnote}{\fnsymbol{footnote}}
\begin{center}
{\LARGE Freeze-free cosmological evolution\\ with a
  non-monotonic internal clock}\\
\vspace{1.5em}
Luis Mart\'{\i}nez,\footnote{e-mail address: {\tt lxm471@psu.edu}}
Martin Bojowald\footnote{e-mail address: {\tt bojowald@psu.edu}}
and Garrett Wendel\footnote{e-mail address: {\tt  gmw5164@psu.edu}}
\\
\vspace{0.5em}
Institute for Gravitation and the Cosmos,\\
The Pennsylvania State
University,\\
104 Davey Lab, University Park, PA 16802, USA\\
\vspace{1.5em}
\end{center}
}

\setcounter{footnote}{0}

\begin{abstract}
  Given the lack of an absolute time parameter in general relativistic
  systems, quantum cosmology often describes the expansion of the universe in
  terms of relational changes between two degrees of freedom, such as matter
  and geometry. However, if clock degrees of freedom (self-)interact
  non-trivially, they in general have turning points where their momenta
  vanish. At and beyond a turning point, the evolution of other degrees of
  freedom is no longer described directly by changes of the clock parameter
  because it stops and then turns back, while time is moving forward. Previous
  attempts to describe quantum evolution relative to a clock with turning
  points have failed and led to frozen evolution in which degrees of freedom
  remain constant while the clock parameter, interpreted directly as a
  substitute for monotonic time, is being pushed beyond its turning
  point. Here, a new method previously used in oscillator systems is applied
  to a tractable cosmological model, given by an isotropic universe with
  spatial curvature and scalar matter. The re-collapsing scale factor presents
  an example of a clock with a single turning point.  The method succeeds in
  defining unitary and freeze-free evolution by unwinding the turning point of
  the clock, introducing an effective monotonic time parameter that is related
  to but not identical with the non-monotonic clock degree of freedom.
  Characteristic new quantum features are found around the turning point,
  based on analytical and numerical calculations.
\end{abstract}

\section{Introduction}

General relativistic systems such as cosmological models are time
reparameterization invariant and therefore lack a physically defined energy
scale. Their canonical description requires a generalization from Hamilton's
equations to constrained evolution, in which both time reparameterizations and
evolution are generated by a single object, the Hamiltonian constraint. A
common method to describe (physically observable) evolution within this
setting and to distinguish it from mere reparamneterizations of a time
coordinate, going back to \cite{GenHamDyn1}, consists in deriving changes of
the dynamical variables with respect to changes of a distinguished one among
them, identified as an internal time parameter or a clock
variable. Similar questions have been analyzed recently in the context of
quantum reference frames
\cite{QuantumRef1,QuantumRef2,QuantumRef3,QuantumRef4,QuantumRef5,QuantumRef6}. 

Classically, the transition from the usual time coordinate $t$ to an internal
clock $\phi$ locally consists in a simple substitution of the local inverse of
a solution $\phi(t)$ for $t$.  However, the local nature of this procedure
implies obstacles at the quantum level. In practice, the procedure has
therefore required the choice of rather special matter systems as candidates
for internal clocks, such as a free, massless scalar field \cite{Blyth}, dust
\cite{BrownKuchar,HusainDust}, or, more generally, systems with purely kinetic
energy and no potential or (self-)interactions. These choices classically
imply monotonic solutions for the relevant fields as functions of coordinate
time, which can be inverted globally. They also have conserved momenta, which
can then serve as simple Hamiltonians that globally generate classical or
quantum evolution with respect to the canonically conjugate variable. While
such models therefore have consistent quantizations, their highly restrictive
and non-fundamental nature means that the physical viability of their
implications should be tested by eliminating the strong underlying
assumptions. An analysis of models with non-monotonic matter solutions or
non-global internal clocks is therefore required, but previous attempts in
quantum cosmology were unable to extend evolution across a turning point of a
non-monotonic clock \cite{WaldTimeModels,PhysEvolBI}: evolution froze because
all degrees of freedom remained constant beyond the turning point of the
clock.

Based on our earlier work on oscillating clocks in quantum mechanics
\cite{Gribov,LocalTime,Period}, we here perform a new analysis of a
cosmological model using a non-global clock variable. The variable we choose
as a clock, given by the scale factor in a closed isotropic model, has a
single turning point and is not oscillating. Our previous methods are seen to
apply nonetheless and show that the resulting dynamics of a scalar field is
consistent. Unlike in previous attempts, the use of an internal clock with
turning points does not lead to freezing cosmological evolution in our
treatment. This outcome relies on the construction of an effective monotonic
time parameter from a non-monotonic fundamental clock. After an introduction
of the classical model in Section~\ref{s:Class}, we present a detailed
construction of global evolution in Section~\ref{s:Global}. Semiclassical
evolution far from the turning point, before as well as after, is described
well by what is expected classically if one uses the same clock for local
evolution. There are new quantum effects around the turning point which we
will derive analytically and confirm by numerical solutions. Some of these
results will be used in a comparison with Dirac observables in
Section~\ref{s:Dirac}.

\section{Classical model}
\label{s:Class}

We begin with the Friedmann equation
\begin{equation}
  \left(\frac{\dot{a}}{a}\right)^2+ \frac{k}{a^2}=\frac{8\pi G}{3} \rho
\end{equation}
for an isotropic model with positive spatial curvature, the dot indicating a
derivative by proper time. The constant $k$ is positive and equals $k=1$ if
the full volume of a spatial 3-sphere is evolved, but it will be convenient to
keep it as a variable constant in our quantum model. In particular, $k$ may be
smaller than one if only a subset of the 3-sphere is considered. While the
choice of this value makes no difference classically (as long as it remains
positive), quantum effects usually depend on the size of the region in which
they are computed, as known from fluctuation energies or Casimir
forces. Similarly, quantum cosmology is sensitive to a change of $k$ by
subdividing the spatial volume as a consequence of infrared renormalization
\cite{Infrared}.

In preparation for our quantization, we first introduce the canonical momentum
\begin{equation}
  p_a=-\frac{3}{4\pi G} a\dot{a}
\end{equation}
and specialize the energy density $\rho$ to the kinetic term
$\rho=\frac{1}{2}p_{\phi}^2/a^6$ of a free, massless scalar $\phi$ with
momentum $p_{\phi}$:
\begin{equation}
  \frac{4\pi G}{3} \left(\frac{p_a}{a^2}\right)^2+ \frac{3k}{4\pi G a^2}=\frac{p_{\phi}^2}{a^6},.
\end{equation}
A first canonical transformation from $(a,p_a)$ to
\begin{equation}
  \tilde{a}=\sqrt{\frac{3}{4\pi G}}\:a\quad\mbox{and}\quad \tilde{p}_a=\sqrt{\frac{4\pi
    G}{3}}\:p_a
\end{equation}
and from $(\phi,p_{\phi})$ to
\begin{equation}
  \tilde{\phi}=\sqrt{\frac{4\pi Gk}{3}}\: \phi\quad\mbox{and}\quad
  \tilde{p}_{\phi}= \sqrt{\frac{3}{4\pi G k}}\: p_{\phi}
\end{equation}
eliminates several constant factors, such that
\begin{equation}
  \left(\frac{\tilde{p}_a}{\sqrt{k}\,\tilde{a}^2}\right)^2+\frac{1}{\tilde{a}^2}=
  \frac{\tilde{p}_{\phi}^2}{\tilde{a}^6}\,.
\end{equation}

A second canonical transformation from $(\tilde{a},\tilde{p}_a)$ to
\begin{equation}
  \alpha=\sqrt{k}\ln\tilde{a}\quad\mbox{and}\quad p_{\alpha}=\frac{\tilde{a}\tilde{p}_a}{\sqrt{k}}
\end{equation}
then implies the constraint
\begin{equation}  \label{qhamiltonian}
  \mathcal{C}=\tilde{p}_{\phi}^{2}-p_{\alpha}^{2}-e^{\gamma\alpha}=0\
\end{equation}
with $\gamma=4/\sqrt{k}\geq 4$. From now on, we will drop the tilde on $\phi$ and $p_{\phi}$
for convenience.

Classically, in terms of some gauge variable $\epsilon$, Hamilton's equations
generated by $\mathcal{C}$ are
\begin{equation}
  \frac{{\rm d}\alpha}{{\rm
      d}\epsilon}=\{\alpha,\mathcal{C}\}=-2p_{\alpha}\approx -2{\rm sgn}(p_{\alpha})
  \sqrt{p_{\phi}^2-e^{\gamma\alpha}} 
\end{equation}
and
\begin{equation}
  \frac{{\rm d}p_{\alpha}}{{\rm d}\epsilon}=\{p_{\alpha},\mathcal{C}\}=\gamma
  e^{\gamma\alpha}\approx 
  \gamma(p_{\phi}^2-p_{\alpha}^2)\,.
\end{equation}
As indicated by the weak equalities, these equations decouple on the
constraint surface $\mathcal{C}=0$, also using the fact that $p_{\phi}$ is
constant thanks to the $\phi$-independence of $\mathcal{C}$. Therefore,
\begin{equation} \label{epssol}
 \alpha(\epsilon)= -\frac{2}{\gamma} \ln\frac{\cosh(\gamma
   |p_{\phi}|\epsilon)}{|p_{\phi}|}\quad,\quad   p_{\alpha}(\epsilon) =
   p_{\phi} \tanh(\gamma p_{\phi}\epsilon) 
\end{equation}
if we eliminate the sole integration constant that remains after imposing the
constraint by choosing $\epsilon$ such that $p_{\alpha}(0)=0$.

For the scalar field we obtain the simple monotonic solution
$\phi(\epsilon)=2p_{\phi}\epsilon+c$ with a constant $c$, as well as constant
$p_{\phi}$.  The model therefore has a global internal time $\phi$, in which
case evolution is generated by the strictly positive Hamiltonian
$-p_{\phi}= \sqrt{p_{\alpha}^2+e^{\gamma\alpha}}$. The equations of motion for
$\phi$-evolution generated by $-p_{\phi}$ can be solved as well, and they
agree with a simple substitution of $p_{\phi}\epsilon=\frac{1}{2}(\phi-c)$ in
(\ref{epssol}). Inverting the resulting $\alpha(\phi)$, we obtain the
double-valued function
\begin{equation} \label{phiclass}
  \phi(\alpha)=c+\frac{2}{\gamma} \cosh^{-1}(|p_{\phi}|e^{-\gamma \alpha/2})\,.
\end{equation}
The branch of the inverse cosh should be switched when $\alpha$ starts
decreasing (if it has been increasing initially) at its turning
point
\begin{equation} \label{alphat}
  \alpha_{\rm t}(p_{\phi})=\frac{2\ln |p_{\phi}|}{\gamma}\,.
\end{equation}

An application of $\alpha$ as an internal clock instead of $\phi$ is therefore
challenging, in particular when the system is to be quantized. We will present
the necessary steps for such a construction in the next section, but first
note that classical evolution with respect to $\alpha$ is locally generated by
the Hamiltonian
\begin{equation} \label{palpha}
  -p_{\alpha}= \pm \sqrt{p_{\phi}^2-e^{\gamma\alpha}}\,.
\end{equation}
Unlike the $\phi$-Hamiltonian, this expression is explicitly clock ($\alpha$)
dependent, and it is not positive definite.  It equals zero at the turning
point (\ref{alphat}) where $\alpha$ has to start decreasing (again, if it has
been increasing initially) for the square root to remain
real. Correspondingly, the sign chosen in (\ref{palpha}) has to be flipped for
$\alpha$ to turn around. The formalism of \cite{Gribov,LocalTime,Period}
allows us to implement this turning point of $\alpha$ even within unitary
quantum evolution determined by an effective time parameter related to (but
not identical with) $\alpha$ as an internal clock.

\section{Global quantum evolution}
\label{s:Global}

Equation~(\ref{palpha}) can formally be quantized to
\begin{equation}\label{Schroalpha}
  i\hbar\frac{\partial\psi(p_{\phi},\alpha)}{\partial\alpha} =
  \pm\sqrt{p_{\phi}^2-e^{\gamma\alpha}}\, \psi(p_{\phi},\alpha)\,.
\end{equation}
However, the Hamiltonian on the right-hand side is not self-adjoint on the
% >>>
kinematical
% <<<
Hilbert space $L^2({\mathbb R}^2,{\rm d}p_{\phi}{\rm d}\alpha)$ and
therefore does not define global evolution with respect to $\alpha$. In order
to ensure that $\alpha$ does not surpass the turning-point value
$\alpha_{\rm t}(p_{\phi})$ given in (\ref{alphat}), we replace $\alpha$ in evolution equations
with an effective time parameter $\tau$ that locally follows the changes of
$\alpha$ but implements the condition that $\alpha$ start decreasing
(with respect to this new $\tau$) right after it reaches the turning
point. These conditions, together with continuity of $\alpha(\tau)$, determine
this relationship as
\begin{equation} \label{parameter}
    \alpha(\tau)=\left\{
        \begin{array}{lc}
            +\tau& \quad\mbox{if }\tau<\alpha_{t}(p_{\phi})   \\
            -\tau+2\alpha_{t}(p_{\phi})&\quad \mbox{if } \tau>\alpha_{t}(p_{\phi})
        \end{array}
    \right.
\end{equation}
% >>>
up to a constant shift. (If a different clock rate is preferred, $\tau$ may be
transformed non-linearly in our final solutions; see Sec.~\ref{s:Gen}.)
% <<<
While the clock variable $\alpha$ is allowed to be non-monotonic around the turning point,
time $\tau$ continues to increase. This crucial step,
clearly distinguishing between clock and time, is sufficient for a consistent
definition of global evolution \cite{Gribov}.
% >>>
In our treatment, time is an effective parameter constructed from a
fundamental clock according to the prescription (\ref{parameter}) that keeps
track of turning points, just  as time labels used in daily life are
constructed from clock readings as well as turning points. Since the time
parameter is not fundamental, we make no attempt to complete (\ref{parameter})
to a canonical transformation, which due to multi-valuedness would require an
extension of the phase space.
% <<<

An important feature of quantum evolution is that a generic state, as a
superposition of $\hat{p}_{\phi}$-eigenstates, has contributions that go
through the turning point at different values of the clock, thanks to the
$p_{\phi}$-dependence of $\alpha_{\rm t}(p_{\phi})$. We will nevertheless
continue to speak of a single turning point because it is unique for a given
$p_{\phi}$, unlike the examples of oscillating clocks studied in \cite{LocalTime,Period}. 

\subsection{Evolution through a turning point}

Using $\tau$ as time, we have to rewrite (\ref{Schroalpha}) in terms of a time
derivative by $\tau$, applying the chain rule and the parameterization (\ref{parameter}):
\begin{equation} 
  i\hbar\frac{\partial\psi(p_{\phi},\tau)}{\partial\tau}=i\hbar\frac{{\rm d}\alpha}{{\rm
      d}\tau}\frac{\partial\psi(p_{\phi},\tau)}{\partial\alpha}
 = \pm \frac{{\rm d}\alpha}{{\rm
     d}\tau}\sqrt{p_{\phi}^2-e^{\gamma\alpha(\tau)}}\, \psi(p_{\phi},\tau)\,. \label{HamS}
\end{equation}
The sign choice has to be made such that $\pm{\rm d}\alpha/{\rm d}\tau=1$,
ensuring stability by a positive $\tau$-Hamiltonian. With this condition,
evolution with respect to $\tau$ is unitary and stable (positive Hamiltonian)
on the Hilbert space $L^2({\mathbb R},{\rm d}p_{\phi})$.

% >>>
The variable $\alpha$ is quantized to an operator only before the constraint
or the evolution equation is imposed (on the kinematical Hilbert space). In
(\ref{HamS}), it has been replacded completely by an effective time parameter
without quantum properties.  Since (\ref{parameter}) is not part of a
canonical transformation, (\ref{HamS}) is not directly constructed from a
quantized momentum of $\tau$, but rather by applying the change rule to the
original equation (\ref{Schroalpha}), which was based on a quantized momentum
of $\alpha$ on the kinematical Hilbert space. This treatment is consistent
with our definition of $\tau$ as an effective time parameter, distinct from
the values of a fundamental clock.  Looking back, equation (\ref{Schroalpha})
presents the first step of our construction in which $\alpha$ is introduced as
the clock variable. If there were now turning points, $\alpha$ would be an
internal time and (\ref{Schroalpha}) would present our evolution equation for
states in the physical Hilbert space $L^2({\mathbb R},{\rm d}p_{\phi})$. Since
there are turning points in our model, we need to perform the second step of
our construction, introducing time by (\ref{parameter}) and formulating
evolution by (\ref{HamS}). Solutions to this equation correspond to physical
states in systems without turning points.
% <<<

The sign choice implies that we should solve the evolution equation separately
for the two cases in (\ref{parameter}), depending on whether $\tau$ is before
or after the turning point $\alpha_{\rm t}(p_{\phi})$ for a given $p_{\phi}$.
% >>>
A state written in the $p_{\phi}$-representation
% <<< A $\hat{p}_{\phi}$-eigenstate $\psi_{p_{\phi}}$ with eigenvalue $p_{\phi}$
evolves as
\begin{equation} \label{solsch}
  \psi_{p_{\phi}}(\tau)=f(p_{\phi})\exp{\Big(-i\Theta(\alpha(\tau),p_{\phi})\Big)}
\end{equation}
with some function $f(p_{\phi})$ for $\tau<\alpha_{\rm t}(p_{\phi})$, where
${\rm d}\alpha/{\rm d}\tau=1$ for the given $p_{\phi}$, with the phase
\begin{equation} \label{phase}
  \Theta(\alpha,p_{\phi})=\frac{2}{\hbar\gamma}\Big(\sqrt{p_{\phi}^{2}-e^{\gamma\alpha}}-
  |p_{\phi}|\tanh^{-1}\sqrt{1-e^{\gamma\alpha}/p_{\phi}^2}\:\Big) \leq 0 \,.
\end{equation}
This function vanishes at the turning point $\alpha_{\rm t}(p_{\phi})$.  Evolving
onwards in the same state with respect to $\tau$ after this value is
reached, $\alpha$ then decreases such that ${\rm d}\alpha/{\rm d}\tau=-1$ in
(\ref{HamS}), and we now choose the minus sign of $\pm$. The corresponding
solution for the wave function is obtained from (\ref{solsch}) by a simple
sign change:
\begin{equation} \label{solsch2}
  \psi_{p_{\phi}}(\tau)=f(p_{\phi})\exp{\Big(i\Theta(\alpha(\tau),p_{\phi})\Big)}\,.
\end{equation}
The phase, now given by $-\Theta(\alpha(\tau),p_{\phi})$, therefore becomes
positive and continues to increase after it goes through zero at the turning
point. For $\alpha\to-\infty$ and using
$\tanh^{-1}(1-x)\sim -\frac{1}{2}\ln(x/2)$ for
$0<x=\frac{1}{2}e^{\gamma\alpha}/p_{\phi}^2\ll1$, we have
$\Theta(\alpha,p_{\phi})\sim \alpha |p_{\phi}|/\hbar$ or
${\rm sgn}({\rm d}\alpha/{\rm d}\tau)\Theta(\alpha(\tau),p_{\phi})\sim \tau
|p_{\phi}|/\hbar$. The combination of (\ref{solsch}) and (\ref{solsch2})
together with (\ref{parameter}) therefore implies that we approximate the
standard linear phase in stationary states
$\psi_{p_{\phi}}(\tau)\propto \exp(-i \tau |p_{\phi}|/\hbar)$ far from the
turning point. There are, however, effects of a non-linear phase around the
turning point implied by the time-dependent Hamiltonian as well as
superposition effects in states that are not eigenstates of $\hat{p}_{\phi}$.

\subsection{Evolving expectation value}

We are interested in analyzing the influence of turning points on the
evolving expectation value of $\hat{\phi}$ in order to show that evolution
does not freeze at a turning point and respects (at least semiclassically) the
monotonic behavior (\ref{phiclass}). To this end, we should use a
suitable normalizable superposition of $\hat{p}_{\phi}$-eigenstates. As usual,
such a superposition is determined by a corresponding function
for the coefficients $f(p_{\phi})$ of these eigenstates. If this function has
support on all $p_{\phi}$, such as a Gaussian state, at any finite time $\tau$
where we may impose an initial state there will be some
$\hat{p}_{\phi}$-eigenstates that have already crossed their turning points,
for which (\ref{solsch2}) should be used, and some which are still approaching
their turning points, for which (\ref{solsch}) should be used. If the initial
time is $\tau_0$ with state $\psi(p_{\phi},\tau_0)=f(p_{\phi})$, the evolved
state for any $\tau\geq\tau_0$ is given by
\begin{equation} \label{psi1}
  \psi(p_{\phi},\tau)= \left\{ \begin{array}{cl}
                                 f(p_{\phi})e^{-i\Theta_0(p_{\phi})}\exp{\Big(+i\Theta(\alpha(\tau),p_{\phi})\Big)}
                                 & \mbox{if }\alpha_{\rm t}(p_{\phi})\leq\tau_0\leq\tau\\
                                 f(p_{\phi})e^{+i\Theta_0(p_{\phi})}\exp{\Big(+i\Theta(\alpha(\tau),p_{\phi})\Big)}
                                 & \mbox{if }\tau_0\leq \alpha_{\rm
                                   t}(p_{\phi})\leq\tau \\
                                 f(p_{\phi})e^{+i\Theta_0(p_{\phi})}\exp{\Big(-i\Theta(\alpha(\tau),p_{\phi})\Big)}
                                 & \mbox{if }\tau_0\leq\tau\leq \alpha_{\rm
                                   t}(p_{\phi})
                               \end{array}\right.
                           \end{equation}
where $\Theta_0(p_{\phi})=\Theta(\alpha(\tau_0),p_{\phi})$. The initial state
determines this constant phase contribution, and sign choices are uniquely
fixed by the relationship between $\tau$ and $\alpha_{\rm t}(p_{\phi})$ as
well as continuity in $\tau$. In particular, when $\tau$ crosses $\alpha_{\rm
  t}(p_{\phi})$ for a given $p_{\phi}$, we move from the second case in
(\ref{psi1}) to the third case. Since $\Theta(\alpha_{\rm
  t}(p_{\phi}),p_{\phi})=0$, continuity requires that the constant phase is
the same in these two cases, without a sign change. The first case in
(\ref{psi1}) requires the opposite sign in the constant phase in order to be
consistent with the initial state $\psi(p_{\phi},\tau_0)=f(p_{\phi})$ when
$\tau_0\not=\alpha_{\rm t}(p_{\phi})$. By changing $\tau$ at fixed $\tau_0$
and $p_{\phi}$, the first case can never be turned into the second or third
one; therefore, there it is not subject to a condition on the phase from
continuity in $\tau$.

Similarly, for $\tau\leq\tau_0$ we have
\begin{equation}
  \psi(p_{\phi},\tau)= \left\{ \begin{array}{cl}
                                 f(p_{\phi})e^{-i\Theta_0(p_{\phi})}\exp{\Big(+i\Theta(\alpha(\tau),p_{\phi})\Big)}
                                 & \mbox{if }\alpha_{\rm t}(p_{\phi})\leq\tau\leq\tau_0\\
                                 f(p_{\phi})e^{-i\Theta_0(p_{\phi})}\exp{\Big(-i\Theta(\alpha(\tau),p_{\phi})\Big)}
                                 & \mbox{if }\tau\leq \alpha_{\rm
                                   t}(p_{\phi})\leq\tau_0\\
                                 f(p_{\phi})e^{+i\Theta_0(p_{\phi})}\exp{\Big(-i\Theta(\alpha(\tau),p_{\phi})\Big)}
                                 & \mbox{if }\tau\leq\tau_0\leq \alpha_{\rm
                                   t}(p_{\phi})
                               \end{array}\right.
\end{equation}
with all sign choices determined by the same conditions as before. In our
analytical examples, we will restrict ourselves to the case of $\tau>\tau_0$
with wave packets that are supported mainly (but not completely) on $p_{\phi}$
such that $\tau_0<\alpha_{\rm t}(p_{\phi})$. The second and third cases in
(\ref{psi1}) will then be sufficient in an approximate analysis.

The expectation value of $\phi$ in such an evolving state is given by
\begin{eqnarray}
  \langle\hat{\phi}\rangle(\tau)&=& i\hbar\langle \psi,{\rm d}\psi/{\rm d}
  p_{\phi}\rangle(\tau)= i\hbar\left\langle \frac{f'(\hat{p}_{\phi})}{f(\hat{p}_{\phi)}}\right\rangle
  + \hbar\left \langle {\rm sgn}({\rm d}\hat{\alpha}/{\rm d}\tau)
  \frac{{\rm d}\Theta(\hat{\alpha}(\tau),\hat{p}_{\phi})}{{\rm
                                    d}p_{\phi}}-\frac{{\rm d}\Theta_0(p_{\phi})}{{\rm d} 
                                    p_{\phi}}\right\rangle \nonumber\\
  &=& i\hbar\left\langle \frac{f'(\hat{p}_{\phi})}{f(\hat{p}_{\phi)}}\right\rangle\\
  &&+ \hbar\left \langle {\rm sgn}({\rm d}\hat{\alpha}/{\rm d}\tau)
  \left(\frac{\partial\Theta(\hat{\alpha}(\tau),\hat{p}_{\phi})}{\partial
      p_{\phi}}+ \frac{\partial\hat{\alpha}}{\partial
      p_{\phi}}\frac{\partial\Theta(\hat{\alpha}(\tau),\hat{p}_{\phi})}{\partial
      \alpha}\right) 
      -\frac{\partial \Theta_0(p_{\phi}))}{\partial 
                                    p_{\phi}} \right\rangle \,.\nonumber
\end{eqnarray}
(According to our assumption that $\tau_0<\alpha_{\rm t}(p_{\phi})$ for most
$p_{\phi}$, the constant phase $\Theta_0(p_{\phi})$ does not depend on
$\alpha_{\rm t}$ within the approximation used.)  The first term is
$\tau$-independent and equals the initial expectation value $\phi_0$ of $\phi$
at $\tau=\tau_0$, where $\psi(p_{\phi},\tau_0)=f(p_{\phi})$. The second term
is evaluated in the same state, but has an explicitly $\tau$-dependent
operator, which we are applying in the $p_{\phi}$-representation. It is
important to note that $\hat{\alpha}(\tau)$ is now an operator because its
classical expression depends on $p_{\phi}$ through $\alpha_{\rm
  t}(p_{\phi})$. The sign change in the phase therefore depends on the
$p_{\phi}$-eigenstate in a superposition given by a general state, and the
$\alpha$-dependence of the phase contributes to the $\hat{\phi}$-expectation
value along with the $p_{\phi}$-dependence.

Using our solution for the phase
$\Theta(\alpha(\tau),p_{\phi})$, we obtain
\begin{eqnarray} \label{expect}
  \langle\hat{\phi}\rangle(\tau)&=&\phi_0+\frac{2}{\gamma}
                                    \left\langle\tanh^{-1}\sqrt{1-e^{\gamma\hat{\alpha}(\tau_0)}/\hat{p}_{\phi}^2}\right\rangle\nonumber\\
  &&
                                   -\frac{2}{\gamma}
  \left\langle{\rm sgn}({\rm d}\hat{\alpha}/{\rm
    d}\tau)
  \tanh^{-1}\sqrt{1-e^{\gamma\hat{\alpha}(\tau)}/\hat{p}_{\phi}^2}
                                    \right\rangle\nonumber\\
&&-\frac{4}{\gamma} \left\langle\theta(-{\rm d}\hat{\alpha}/{\rm
    d}\tau) \sqrt{1-e^{\gamma\hat{\alpha}(\tau)}/\hat{p}_{\phi}^2} \right\rangle
\end{eqnarray}
with the Heaviside step function $\theta(x)$ from
\begin{equation} \label{expectcoeff}
  \frac{\partial\alpha(\tau)}{\partial p_{\phi}}=2 \theta(-{\rm d}\alpha/{\rm
    d}\tau) \frac{{\rm d}\alpha_{\rm t}}{{\rm d}p_{\phi}}= \frac{4}{\gamma
    |p_{\phi}|} \theta(-{\rm d}\alpha/{\rm 
    d}\tau) \,.
\end{equation}
combining (\ref{parameter}) and (\ref{alphat}).  The $\tau$-dependent parts
are continuous in $\tau$ in spite of the sign and the Heaviside function
because the latter are multiplied by functions that vanish at the step
where the sign of ${\rm d}\alpha/{\rm d}\tau$ changes.

The functional dependence of the third term in (\ref{expect}) can be seen to equal
(\ref{phiclass}), using $\tanh(x)=\sqrt{\cosh^2(x)-1}/\cosh(x)$, but now the
branch of $\cosh^{-1}$ is explicitly determined by
${\rm sgn}({\rm d}\alpha/{\rm d}\tau)$ for each $p_{\phi}$-eigenstate in a
superposition. The fourth contribution to the expectation value is entirely
determined by the turning points and has no classical analog. The inverse tanh
in the third term implies that the fourth term is relevant only during a time
interval when the majority of the $\hat{p}_{\phi}$-eigenstates are crossing
the turning point.

%>>>2
\subsection{General features of the method}
\label{s:Gen}

Before we continue with a detailed analysis of evolution in our specific
model, we briefly discuss several aspects related to the applicability
of our general method and its time parameterization.

The specific example considered here describes a simple cosmological model in
which most equations can be solved analytically. A large class of
generalizations is available within the same method, but there are also models
that require further developments that are still in progress. Looking at the
constraint, it may be generalized in two ways, by changing the clock
Hamiltonian or the Hamiltonian of the system interpreted as evolving with
respect to the clock.

If we first fix the clock Hamiltonian, it is easy to see
that our methods can be applied to any system Hamiltonian $H(\phi,p_{\phi}$
such that the initial evolution equation (\ref{Schroalpha}) is replaced by
\begin{equation}
  i\hbar\frac{\partial|\psi\rangle(\alpha)}{\partial\alpha} =
  \pm\sqrt{\hat{H}-e^{\gamma\alpha}}\, |\psi\rangle(\alpha)\,.
\end{equation}
In this general case, it is not convenient to work in the
$p_{\phi}$-representation, but our equations can easily be adjusted if we work
in the representation spanned by eigenstates of $\hat{H}$. Given the positive
clock Hamiltonian and the constraint, we can restrict the spectrum of
$\hat{H}$ to its positive part. Assuming the corresponding eigenstates are
labeled by some number $k$ (and additional labels in case of degeneracies in
the positive part of the spectrum), which may be discrete or continuous, and
the $\hat{H}$-eigenvalues are $E_k\geq 0$, the evolution equation
\begin{equation}
  i\hbar\frac{\partial\psi_k(\alpha)}{\partial\alpha} =
  \pm\sqrt{E_k-e^{\gamma\alpha}}\, \psi_k(\alpha)
\end{equation}
can be solved simply by replacing $|p_{\phi}|$ with $\sqrt{E_k}$ in our
previous phase.

The same substitution can be used in the globally evolved wave
function, given for $\tau\geq\tau_0$ by
\begin{equation} \label{psi1gen}
  \psi_k(\tau)= \left\{ \begin{array}{cl}
                                 f(k)e^{-i\Theta_0(k)}\exp{\Big(+i\Theta(\alpha(\tau),k)\Big)}
                                 & \mbox{if }\alpha_{\rm t}(k)\leq\tau_0\leq\tau\\
                                 f(k)e^{+i\Theta_0(k)}\exp{\Big(+i\Theta(\alpha(\tau),k)\Big)}
                                 & \mbox{if }\tau_0\leq \alpha_{\rm
                                   t}(k)\leq\tau \\
                                 f(k)e^{+i\Theta_0(k)}\exp{\Big(-i\Theta(\alpha(\tau),k)\Big)}
                                 & \mbox{if }\tau_0\leq\tau\leq \alpha_{\rm
                                   t}(k)
                               \end{array}\right.
                           \end{equation}
where $\alpha_{\rm t}(k)=\gamma^{-1}\ln(E_k)$, and a corresponding version for
$\tau<\tau_0$. Since $|\psi_k(\tau)|^2=|f(k)|^2$ for all $\tau$, normalization
of this state in the $\hat{H}$-representation is preserved by evolution. The
same property is then true in any other representation, where the evolving
state is
\begin{equation}
|\psi\rangle(\tau) = \int_k \psi_k(\tau) |k\rangle {\rm d}k
\end{equation}
if $|k\rangle$ are the $\hat{H}$-eigenstates in the desired representation. In
this expression, integration over $k$ (or summation in the discrete case)
at fixed $\tau$ takes into account the non-trivial $k$-dependence of the phase in
(\ref{psi1gen}) and may therefore be more complicated than the corresponding
transformation in a model with an absolute time. Nevertheless, 
$\tau$-independence of normalization follows from general properties, in particular
the fact that the $\hat{H}$-eigenbasis is orthonormal for self-adjoint
$\hat{H}$ together with preserved normalization in the
$\hat{H}$-representation based on (\ref{psi1gen}).

It is also possible to apply our methods to a different clock Hamiltonian. We
will not go into details here because doing so would require a new
parameterization of $\alpha(\tau)$, possibly with multiple turning points. The
methods used here can easily be adjusted to other clock potentials if there is
still only one turning-point value, $\alpha_{\rm t}$. The case of two
turning-point values of the form $\pm\alpha_{\rm t}$ has been discussed in
detail in the clock model of \cite{Period,LocalTime}.

Further generalizations that include direct interaction terms between clock
and system, such as a constraint
$C=p_{\alpha}^2+ I(\alpha,\phi) - H(\phi,p_{\phi})$ with an interaction term
$I(\alpha,\phi)$ that depends on both $\alpha$ and $\phi$, remain
challenging. In this case, since $\{I(\alpha,\phi),H(\phi,p_{\phi})\}\not=0$,
we are not able to diagonalize the operators $\hat{I}$ and $\hat{H}$
simultaneously. The $\hat{H}$-representation then does not sufficiently
simplify the dynamics, and without analytical solutions, it is more difficult
to implement suitable phase changes at turning points. Preliminary
investigations suggest that at least a numerical treatment is possible in
principle, but slowed down by the requirement to transform back and forth
between the $\hat{H}$ and the $\hat{I}$-representations. A generalization of
our methods to systems with direct system-clock interactions would certainly
be important for cosmological models.

Within a specific model such as the one used here, one may be interested in
considering different time parameterizations in which the rate of change of
$\tau$ does not agree with the rate of change of the clock degree of freedom,
$\alpha$. For instance, in our cosmological model, $\alpha$ is the logarithmic
scale factor, but one may want to write evolution with respect to proper time,
given by a different function of the scale factor, depending on the model. Our
method allows for such reparameterizations. Instead of (\ref{parameter}), we
may use
\begin{equation} \label{parametergen}
    \alpha(\tau)=\left\{
        \begin{array}{lc}
            +R(\tau)& \quad\mbox{if }R(\tau)<\alpha_{t}(p_{\phi})   \\
            -R(\tau)+2\alpha_{t}(p_{\phi})&\quad \mbox{if } R(\tau)>\alpha_{t}(p_{\phi})
        \end{array}
    \right.
\end{equation}
with a monotonic reparameterization function $R(\tau)$. The reparameterization
function is not applied to $\alpha_{\rm t}(p_{\phi})$ in the branch conditions
and in the second line of (\ref{parametergen}) because these terms are
determined by $\alpha$ or $R(\tau)$ (but not $\tau$) reaching the turning
point. Therefore, derivative operators acting on the $p_{\phi}$-dependence of
the wave function through $\alpha_{\rm t}$, such as (\ref{expect}), and in
particular the coefficient in (\ref{expectcoeff}), are unchanged by the
reparameterization. The only effect of the reparameterization is to
shift the location of turning points in terms of $\tau$ (through ${\rm sgn}({\rm
  d}\hat{\alpha}/{\rm d}\tau$ and $\theta(-{\rm d}\hat{\alpha}/{\rm d}\tau)$
in (\ref{expect})), and to function  as a standard reparameterization of time
away from turning points.

\subsection{Shifts around the turning point}

The functional form of (\ref{expect}) suggests that there may be significant
quantum effects around the turning point, but at much earlier or later times
we have nearly classical behavior provided the initial state is sufficiently
semiclassical. A useful way to express quantum effects is by computing
additional shifts they imply in the asymptotic behavior of $\phi(\tau)$ for
$\tau\to\pm\infty$ in addition to the classical shift. For $\alpha\to-\infty$,
and therefore for both $\tau\to-\infty$ and $\tau\to\infty$, we have
\begin{equation}
  \frac{{\rm
  d}\phi}{{\rm d}\tau}=-{\rm sgn}({\rm d}\alpha/{\rm
d}\tau)\frac{p_{\phi}}{p_{\alpha}}=\frac{|p_{\phi}|}{\sqrt{p_{\phi}^2-e^{\gamma\alpha}}}
  \to 1
\end{equation}
such that $\phi(\tau)$ asymptotically approaches a straight line at an
angle of $45^{\circ}$. However, the behavior around the turning point is not
linear, which implies a constant shift in the $\phi$-direction between the
asymptotic past and the asymptotic future. This shift is sensitive to quantum
effects around the turning point and can therefore be used to quantify them.

We first compute the classical shift as a reference basis. We can introduce in
the classical solution $\phi(\alpha)$, given in equation~(\ref{phiclass}), the
same parameterization $\alpha(\tau)$ given in (\ref{parameter}) as used in the
unwinding of $\alpha$ as a quantum clock. Classically, the resulting
$\phi(\tau)$ can be viewed as a reparameterization of the gauge orbit
$\phi(\epsilon)$ with a non-linear transformation $\tau(\epsilon)$ such that
${\rm d}\alpha/{\rm d}\epsilon=\pm {\rm d}\tau/{\rm d}\epsilon$. If the
initial value $\phi=\phi_0$ is chosen at some
$\tau_0<\alpha_{\rm t}(p_{\phi})$ for a given $p_{\phi}$, continuity of
$\phi(\tau)$ requires that
\begin{eqnarray}
  \phi(\tau)= \left\{\begin{array}{cl} \phi_0+2\gamma^{-1}
                       \tanh^{-1}\sqrt{1-e^{\gamma\tau_0}/p_{\phi}^2}
                       -2\gamma^{-1} \tanh^{-1}\sqrt{1-e^{\gamma\alpha(\tau)}/p_{\phi}^2}&
                                                                        \mbox{if
                                                                        }\tau<\alpha_{\rm
                                                                        t}(p_{\phi})\\
                       \phi_0+2\gamma^{-1}
                       \tanh^{-1}\sqrt{1-e^{\gamma\tau_0}/p_{\phi}^2} 
                       +2\gamma^{-1} \tanh^{-1}\sqrt{1-e^{\gamma\alpha(\tau)}/p_{\phi}^2}&
                                                                        \mbox{if
                                                                        }\tau>\alpha_{\rm
                                                                        t}(p_{\phi})\end{array}\right. \,. 
\end{eqnarray}
This function is consistent with (\ref{epssol}) for all $\tau$ if we use
\begin{equation}
  \phi(\tau(\epsilon))=2p_{\phi}\epsilon+ \phi_0+ 2\gamma^{-1}
  \tanh^{-1}\sqrt{1-e^{\gamma\tau_0}/p_{\phi}^2}
\end{equation}
to reparameterize from $\tau$
to $\epsilon$. Using again $\tanh^{-1}(1-x)\sim -\frac{1}{2}\ln(x/2)$ as
already applied in our analysis of the phase, we have
\begin{equation} \label{expandtanh}
  -\frac{2}{\gamma} {\rm sgn}({\rm d}\alpha/{\rm d}\tau)
  \tanh^{-1}\sqrt{1-e^{\gamma\alpha}/p_{\phi}^2} \sim {\rm sgn}({\rm
    d}\alpha/{\rm d}\tau)  \alpha - \frac{2}{\gamma} {\rm sgn}({\rm
    d}\alpha/{\rm d}\tau)\ln(2|p_{\phi}|)\,. 
\end{equation}
If we simply combine the two $\alpha$-branches, their asymptotic linear curves
are separated by a $\phi$-shift of $4\gamma^{-1} \ln(2|p_{\phi}|)$. However, the
parameterization in terms of $\tau$ changes this result because of an
additional shift of $\tau$ by
$2\alpha_{\rm t}(p_{\phi})=4\gamma^{-1}\ln(|p_{\phi}|)$ according to
(\ref{parameter}):
\begin{eqnarray}
&&  {\rm sgn}({\rm
    d}\alpha/{\rm d}\tau) \alpha(\tau) - \frac{1}{\gamma} {\rm sgn}({\rm
    d}\alpha/{\rm d}\tau)\ln(2p_{\phi}^2)\nonumber\\&=& \left\{\begin{array}{cl} \tau-
                                                    2\gamma^{-1}
                                                    \ln(2|p_{\phi}|) &
                                                                       \mbox{if
                                                                       }\tau<\alpha_{\rm
                                                                       t}\\
                                                    \tau-2\alpha_{\rm
                                                    t}(p_{\phi})+
                                                    2\gamma^{-1}\ln(2|p_{\phi}|)
                                                                     &
                                                                       \mbox{if
                                                                       }\tau>\alpha_{\rm
                                                                       t} \end{array}\right.\nonumber\\
                                                                 &=& \left\{\begin{array}{cl} \tau-
                                                    2\gamma^{-1}
                                                    \ln(2|p_{\phi}|) &
                                                                       \mbox{if
                                                                       }\tau<\alpha_{\rm
                                                                       t}\\
                                                    \tau-
                                                    2\gamma^{-1}\ln(2|p_{\phi}|) +4\gamma^{-1}\ln2
                                                                     &
                                                                       \mbox{if
                                                                       }\tau>\alpha_{\rm
                                                                       t} \end{array}\right.
\end{eqnarray}
The classical shift is therefore independent of $p_{\phi}$ and is given by
\begin{equation}
  \Delta\phi_{\rm classical}= \frac{4}{\gamma}\ln2\,.
\end{equation}

The same shift appears in the quantum case, but there are additional
contributions as well.
For $\tau\to-\infty$, the fourth term in (\ref{expect}) can be
ignored because the majority of $\hat{p}_{\phi}$-eigenstates still has to
cross their turning points. The third term then guarantees nearly classical
behavior.  For $\tau\to\infty$ when $\alpha(\tau)\to-\infty$ and most
eigenstates have gone through their turning points, the fourth term only
implies a constant negative shift of $\langle\hat{\phi}\rangle(\tau)$ by
\begin{equation} \label{Deltaphi}
  \Delta\phi_1=-\frac{4}{\gamma}
\end{equation}
while the inverse tanh in the second term continues to increase with linear
asymptotic $\tau$-dependence  as
$\sqrt{1-e^{\gamma\alpha(\tau)}/p_{\phi}^2}$ approaches one.
The full quantum shift equals
\begin{equation}
  \Delta\phi=\Delta\phi_{\rm classical}+\Delta\phi_1
  = \Delta\phi_{\rm classical}
  -\frac{4}{\gamma}
\end{equation}
provided the initial state is posed sufficiently far ahead of the turning
point. For $\gamma=4$, the shifts simplify to $\Delta\phi_{\rm
  classical}=\ln(2)$ and $\Delta\phi=\Delta\phi_{\rm classical}-1$.

As a function of $\tau$, the classical-type contribution to (\ref{expect}),
depending on $\tanh^{-1}$, provides a monotonic asymptotic contribution to
$\langle\hat{\phi}\rangle(\tau)$: While
$\sqrt{1-e^{\gamma\alpha(\tau)}/p_{\phi}^2}$ has a local minimum of zero at
$\tau=\alpha_{\rm t}(p_{\phi})$, and so does
$\tanh^{-1}\sqrt{1-e^{\gamma\alpha(\tau)}/p_{\phi}^2}$, multiplying this
expression with $-{\rm sgn}({\rm d}\alpha/{\rm d}\tau)$ turns it
into a monotonically increasing function of $\tau$. The $\tau$-dependent
expectation value of monotonic functions of $\tau$ is also monotonic. Since
the $\tanh^{-1}$ is dominant well before and after the turning point, the
classical monotonic behavior of $\phi(\tau)$ is maintained in this regime.

The last contribution to (\ref{expect}) does not respect this behavior because
it subtracts an increasing function after the turning point, thanks to the
factor of $\theta(-{\rm d}\alpha/{\rm d}\tau)$. Since this term approaches a
constant at late times, it can change the monotonic behavior of $\phi(\tau)$
only around the turning point, the more so far larger $p_{\phi}$ because the
slope of the subtracted square root is then larger close to the turning
point. The quantum shift $\Delta\phi_1$ is implied by the same term, but it is
independent of $p_{\phi}$ because it refers to the asymptotic value.

While there may therefore be noticeable quantum effects around the turning
point of $\alpha$, these observations demonstrate that our quantum evolution
does not freeze there and has the correct asymptotic behavior, before and
after the turning point. The eigenstate dependence of the turning points makes
these features more involved than in standard quantum mechanics. We now turn
to numerics in order to show specific examples of evolving states.

\subsection{Numerical treatment}

For a numerical analysis of equation~(\ref{expect}) we have to choose an
initial state, which we take to be a Gaussian
\begin{equation}
  \psi_{0}(\phi)=(2\pi d^2)^{-1/4}\exp{\Big(\frac{i\phi p_{\phi}}{\hbar}-\frac{(\phi-\phi_{0})^{2}}{4d^2}\Big)}
\end{equation}
centered at $(\phi_0,p_{\phi})$.  Numerically, we restrict the $\phi$-range to
be a finite interval between $-L$ and $L$ for some $L$ sufficiently large to
contain a given stretch of evolving $\phi(\tau)$. In addition, we express the
evolved state as a truncated sum of $\hat{p}_{\phi}$-eigenstates,
\begin{equation}\label{expansion1}
  \psi(\tau,\phi)=\sum_{n=-N}^{+N} c_{n} \exp{\Big(-i {\rm
      sgn}({\rm d}\alpha(\tau,p_{\phi}^{(n)})/{\rm
      d}\tau)\Theta(\alpha(\tau,p_{\phi}^{(n)}),p_{\phi}^{(n)})\Big)}\exp{(i\phi
    p_{\phi}^{(n)}/\hbar)} \,,
\end{equation}
where $p_{\phi}^{(n)}=n\pi/L$. In our examples, the normalization of $\psi(\tau,\phi)$
is approximately conserved within a relative accuracy of $10^{-4}$ for
$N/L>10$.

Since the turning point (\ref{alphat}) depends on $p_{\phi}$, the truncated
superposition (\ref{expansion1}) experiences a turning point at $2N+1$
different values of $\tau$. A non-truncated state supported on the infinite
set of momenta $p_{\phi}$ will never completely cross all the turning points
since at any finite $\tau$ there will be terms in the superposition of
$\hat{p}_{\phi}$-eigenstates that have yet to encounter their time of the
turning point. However, these terms contribute less and less as the wave
function approaches zero for large $|p_{\phi}|$. The finite truncation is
therefore expected to be reliable.

\begin{figure}
\begin{center}
        \includegraphics[width=18cm]{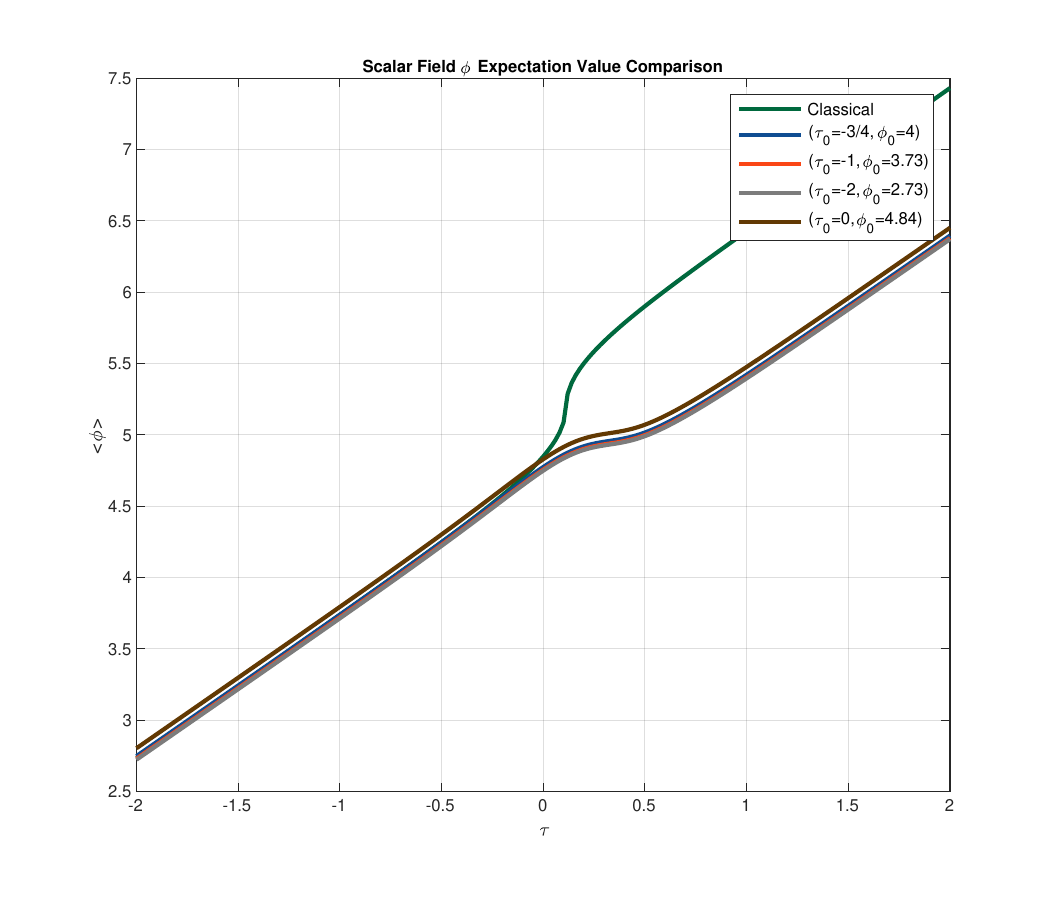}
        \caption{Expectation value of $\phi$ as a function of $\tau$ for
          different initial times $\tau_0$ at which the state is assumed to be
          Gaussian, centered at $\phi_0(\tau_0)$ following the classical
          curve. The differences between the future and past asymptotic
          behaviors in the classical and quantum case, respectively, are
          consistenty with the values
          $\Delta\phi_{\rm classical}\approx 0.7$,
          $\Delta\phi=\Delta\phi_{\rm classical}-1\approx-0.3$ implied by our
          formulas with the value $\gamma=4$. The momentum in this case equals
          $p_{\phi}=1.25$} 
    \label{fig:phitau}
\end{center}
\end{figure}

Examples of the expectation value $\langle\hat{\phi}\rangle(\tau)$ for different
values of the initial time $\tau_0$ where the state is Gaussian are shown in
Fig.~\ref{fig:phitau}. The initial time does not matter much, even if it is
set close to the main turning point. The only changes in the resulting curves
are implied by a slightly different initial $\phi_0(\tau_0)$, which was chosen
to follow the classical behavior.  Confirming our analytical results based on
equation~(\ref{expect}), the field expectation value retains its monotonic
behavior in asymptotic regimes and, for the specific values of $\gamma$ and
$p_{\phi}$ chosen in this figure, even while our local clock $\alpha$ goes
through its turning point. The slope of $\langle\hat{\phi}\rangle(\tau)$
asymptotically approaches the value one if the state remains
semiclassical because for $\alpha\to-\infty$,
\begin{eqnarray}
  \frac{{\rm d}\phi}{{\rm d}\tau}&=& {\rm sgn}({\rm d}\alpha/{\rm d}\tau)
  \frac{{\rm d}\phi/{\rm d}\epsilon}{{\rm 
      d}\alpha/{\rm d}\epsilon}= -{\rm sgn}({\rm d}\alpha/{\rm
                                     d}\tau)\frac{p_{\phi}}{p_{\alpha}}\nonumber\\
  &=&
  \frac{1}{\sqrt{1- e^{\gamma\alpha}/p_{\phi}^2}} \to 1
\end{eqnarray}
is implied by the classical constraint.

\begin{figure}
\begin{center}
        \includegraphics[width=12cm]{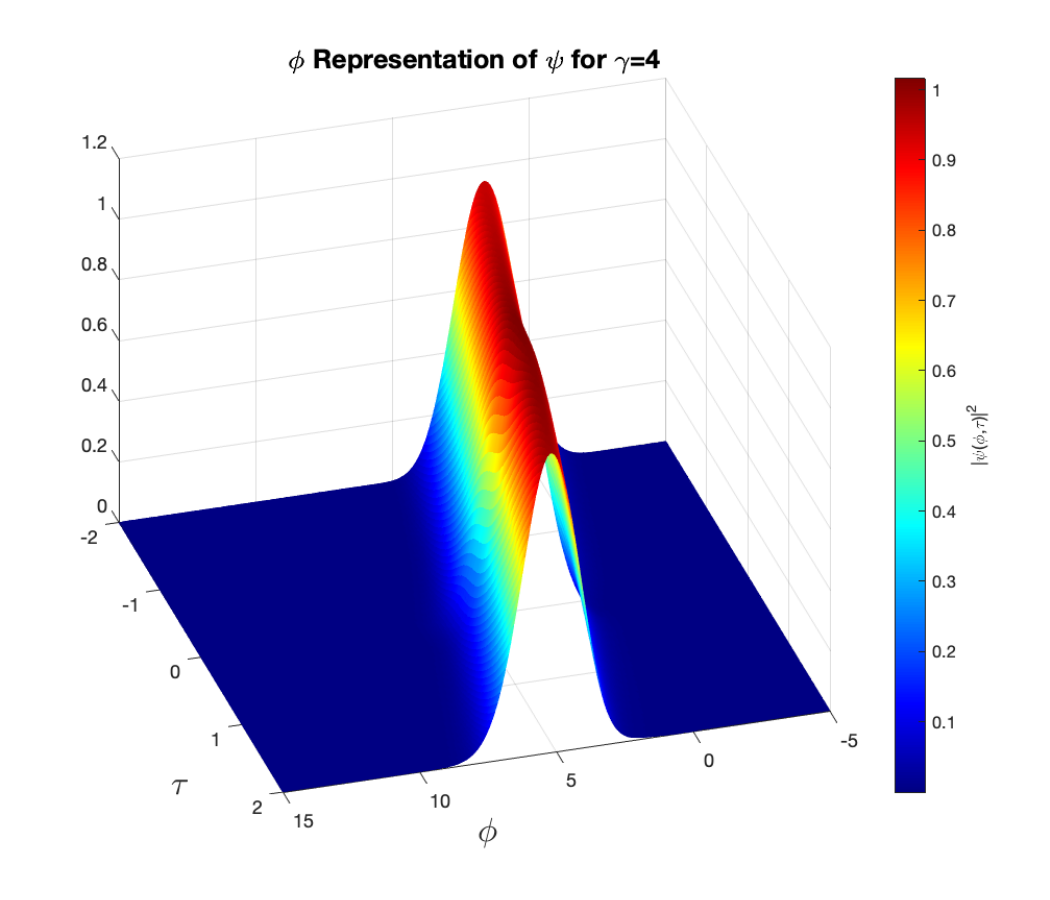}
        \includegraphics[width=12cm]{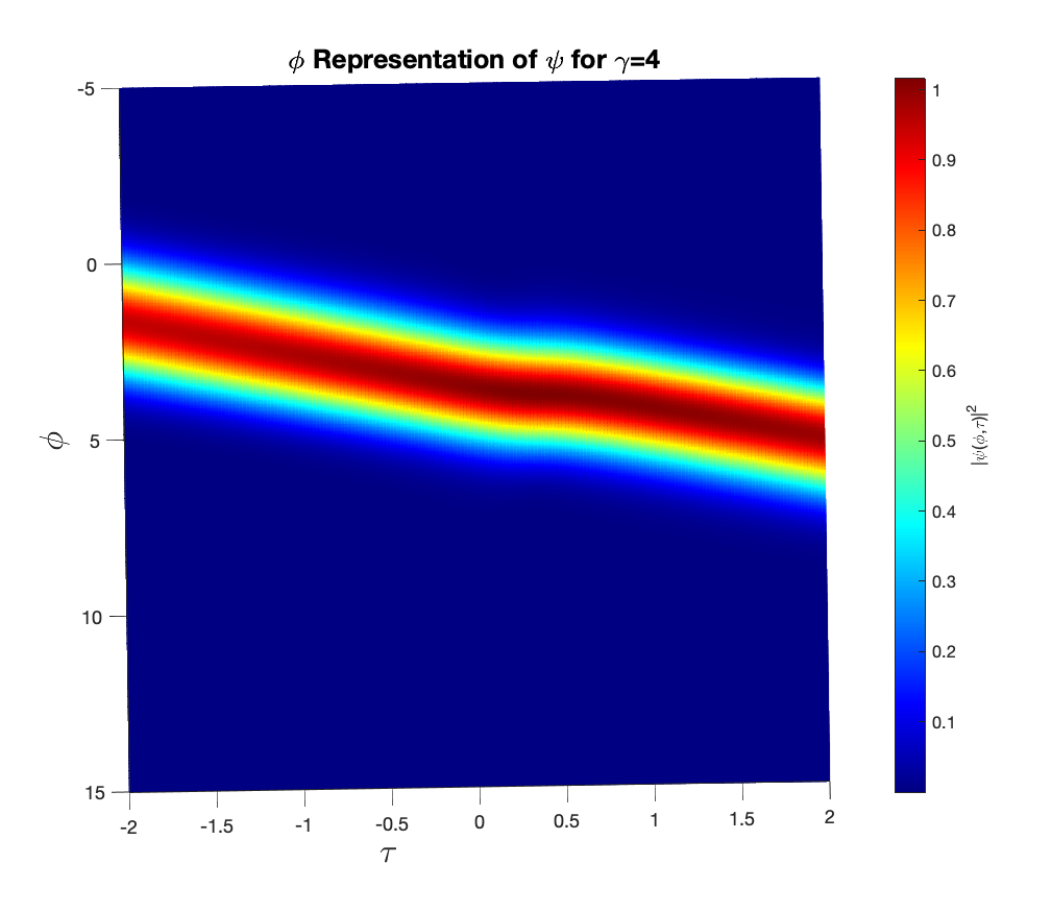}
    \caption{Evolution of the probability density for the Gaussian wavepacket
      under the constraint Hamiltonian for $\gamma=4$ and $p_{\phi}=1.25$.}
    \label{fig:density1}
    \end{center}
\end{figure}

As further support of freeze-free evolution, we show the probability density
of $\hat{\phi}$ in the evolved state in Fig.~\ref{fig:density1}. Our initial
state is Gaussian, but it does not strictly keep this form because the
$\alpha$-Hamiltonian is not harmonic. Most of the turning points happen in the
region close to $\tau=0$ in this case where the magnitude
$|\psi(\phi,\tau)|^2$ is largest. The wave function continues to evolve as
expected while going through the turning point of $\alpha$ for several
$p_{\phi}$-eigenstates.

\begin{figure}
\begin{center}
        \includegraphics[width=17cm]{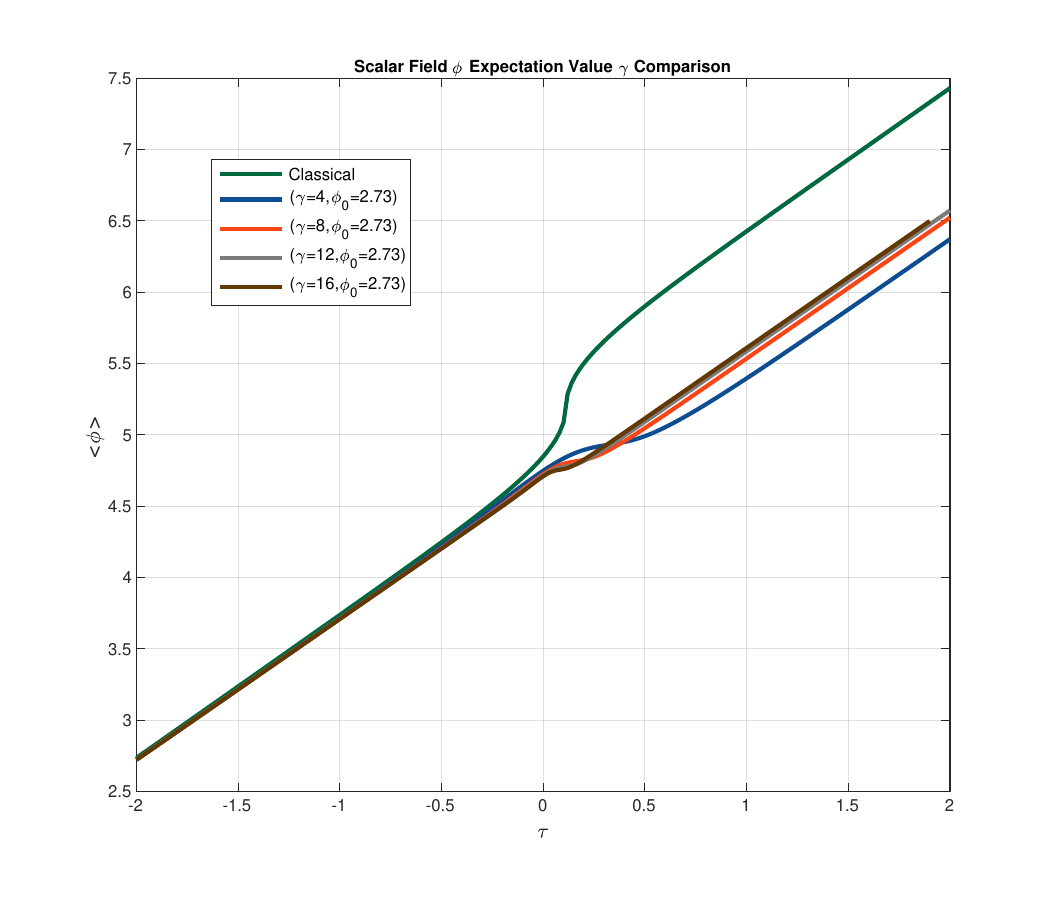}
        \caption{Expectation value of $\phi$ as a function of $\tau$ for three
          values of $\gamma\geq4$. Each curve corresponds to the same Gaussian
          initial state ($d=1$)at the main turning point
          $\tau=2\gamma^{-1}\ln |p_{\phi}|$, where $p_{\phi}=1.2 $ has been chosen
          for the numerics. } 
    \label{fig:phitaugamma}
  \end{center}
\end{figure}

The dependence on $\gamma$ is illustrated in Fig.~\ref{fig:phitaugamma}. As
expected from our analytical expression for the quantum shift, the future
asymptotic behavior of the quantum curve is closer to the classical curve for
larger $\gamma$, but, in contrast to the classical solution, it is always
below an extension the asymptotic past to the future as a straight line.

\begin{figure}
\begin{center}
        \includegraphics[width=17cm]{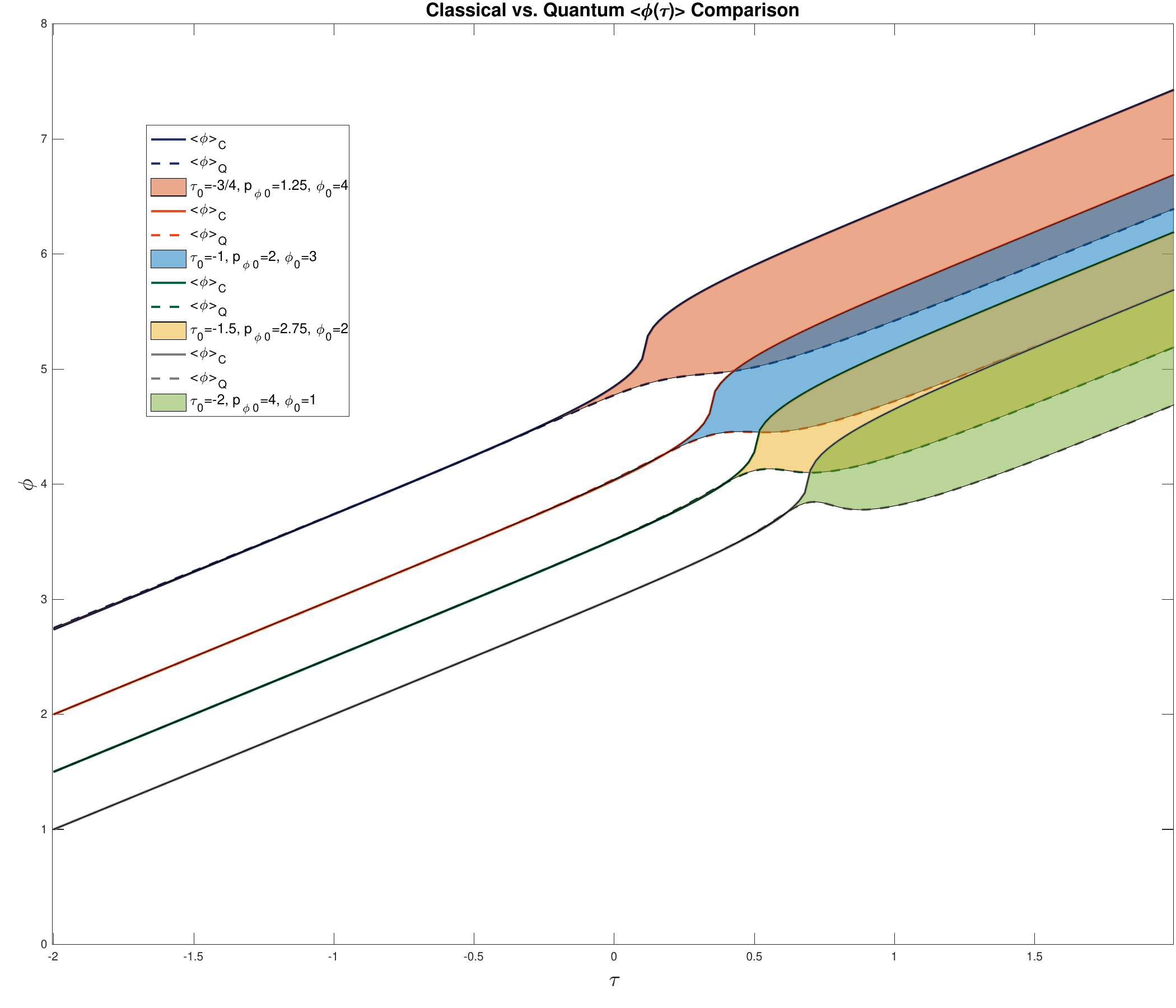}
        \caption{Expectation value of $\phi$ as a function of $\tau$ for four
          values of $p_{\phi}$. For the sake of clarity, the curves have been
          plotted with different initial $\phi_0(\tau_0)$ in order to shift
          them apart. While the asymptotic behavior does not depend on
          $p_{\phi}$, the range around the turning point where the transition
          between the asymptotic past and future happens shows a new
          feature for larger $p_{\phi}$: The expectation value
          $\langle\hat{\phi}\langle(\tau)$ is no longer monotonic for
          sufficiently large values.} 
    \label{fig:phitaup}
\end{center}
\end{figure}

A new feature is shown in the dependence on $p_{\phi}$ in
Fig.~\ref{fig:phitaup}. The asymptotic parts are not affected by changing
$p_{\phi}$, but the behavior around the turning point does change noticeably
when $p_{\phi}$ is increased from the value $p_{\phi}=1.25$ used in the
previous plots. In particular, the expectation value
$\langle\hat{\phi}\rangle(\tau)$ is no longer monotonic in a small range
around the turning point. A closer look at the curves reveals that this
feature is a consequence of a property seen in the analytical expression
(\ref{expect}): Starting in the asymptotic past, the expectation value follows
the classical curve longer for larger $p_{\phi}$, but then approaches the
shifted future behavior more quickly. This behavior is implied by the
appearance of $p_{\phi}$ in $\sqrt{1-e^{\gamma\alpha}/p_{\phi}^2}$, which
leads to a more abrupt transition for larger $p_{\phi}$ and makes quantum
effects of turning points more noticeable. The numerical solutions show that a
sufficiently rapid transition can lead to a non-monotonic
$\langle\hat{\phi}\rangle(\tau)$ in a small interval around the turning point.

\section{Comparison with Dirac quantization}
\label{s:Dirac}

Our classical constraint
$\mathcal{C}=p_{\phi}^2-p_{\alpha}^2-e^{\gamma\alpha}$ can easily be turned
into an operator on the kinematical Hilbert space
$L^2(\mathbb{R}^2,{\rm d}\alpha{\rm d}\phi)$. According to standard Dirac
quantization, the physical Hilbert space on which the constraint is solved
would then be a suitable Hilbert-space completion of (generalized) states
$\psi$ satisfying
\begin{equation} \label{Cpsi}
  \hat{\mathcal{C}}\psi=( \hat{p}_{\phi}^2-\hat{p}_{\alpha}^2-e^{\gamma\hat{\alpha}})\psi=0\,.
\end{equation}

\subsection{Observables and double-valuedness}

A common way \cite{Blyth,GenRepIn} to introduce a physical inner product
interprets (\ref{Cpsi}) in the $(\alpha,\phi)$-representation
$\psi(\alpha,\phi)$ as a Klein--Gordon equation and uses the
bilinear form
\begin{equation} \label{KGform}
 (\psi_1,\psi_2)=  i \int \left( \psi_1^* \frac{\partial\psi_2}{\partial\phi}-
   \frac{\partial\psi_1^*}{\partial\phi} \psi_2\right){\rm d}\alpha
\end{equation}
conserved in evolution with respect to $\phi$. In order to
obtain a positive definite inner product, one restricts the solution space to
positive-frequency solutions with positive eigenvalues of $\hat{p}_{\phi}$, or
combines positive-frequency solutions with the inner product $(\psi_1,\psi_2)$
and negative-frequency solutions with the inner product
$-(\psi_1,\psi_2)$.

Irrespective of the specific choice, the procedure makes use of the
factorization
\begin{equation} \label{Cpm}
  \hat{\mathcal{C}}= \hat{\mathcal{C}}_- \hat{\mathcal{C}}_+=
  \left(\hat{p}_{\phi}-\sqrt{\hat{p}_{\alpha}^2+e^{\gamma\hat{\alpha}}}\right) 
  \left(\hat{p}_{\phi}+\sqrt{\hat{p}_{\alpha}^2+e^{\gamma\hat{\alpha}}}\right)
\end{equation}
into two commuting factors. Both positive-frequency solutions, solving
$\hat{\mathcal{C}}_-\psi=0$, and negative-frequency solutions, solving
$\hat{\mathcal{C}}_+\psi=0$, are therefore part of the solution space of the
constraint. The two types of states can then be interpreted as evolving with
respect to $\phi$, such that (\ref{KGform}) together with the correct sign
choice provides an inner product given by integration over $\alpha$ at some
fixed $\phi$. The value chosen for $\phi$ does not matter because the inner
product is conserved in $\phi$.

Dirac observables provide physically motivated operators acting on the physical Hilbert
space. They are defined as expressions that (classically) have vanishing
Poisson brackets with the constraint or (in a quantization) commute with the
constraint operator. The classical version of such an expression is conserved
by the gauge flow generated by the constraint on phase space, and the quantum
version maps the solution space of the constraint to itself. A simple example
in the present case is the conserved momentum $p_{\phi}$, which locally has a
canonically conjugate Dirac observable
\begin{equation}
  \Phi=\phi-\frac{2}{\gamma} \cosh^{-1}(|p_{\phi}|e^{-\gamma\alpha/2})\,.
\end{equation}

The conserved nature of $\Phi$ can be interpreted as describing evolution of
$\alpha$ relative to $\phi$, such that $\Phi$ remains constant.  The
double-valuedness of the inverse cosh implies that $\Phi$ is only locally
defined on phase space.  Once $\alpha$ reaches a turning point
$\alpha_{\rm t}(p_{\phi})$, the other branch of the inverse cosh should be
followed in order to have agreement between this relational evolution
$\alpha(\phi)$ and the gauge orbits given by $\alpha(\epsilon)$ and
$\phi(\epsilon)$ discussed in Section~\ref{s:Class}. (A global Dirac
observable other than $p_{\phi}$ is given by
\begin{equation}
  D=e^{-\gamma\alpha/2} \left(p_{\phi}\cosh(\gamma\phi/2)-
    p_{\alpha}\sinh(\gamma\phi/2)\right)\,.
\end{equation}
It cannot be solved globally for $\phi(\alpha)$.)

\subsection{Disambiguations}

The double-valuedness of $\Phi$ means that this expression does not have a
straightforward quantization. An additional choice is required that specifies
how the two values are assigned to different states. For instance, the
classical expression $\pm\sqrt{|\alpha|}$, as a simpler version of
$\cosh^{-1}(|p_{\phi}|e^{-\gamma\alpha/2})$, may be disambiguated by assigning
the plus choice to positive values of $\alpha$ and the minus choice to
negative values of $\alpha$, such that
$\pm\sqrt{|\alpha|}={\rm sgn}(\alpha)\sqrt{|\alpha|}$. In a quantization, this
choice can be implemented by using the same disambiguation on the spectral
decomposition of $\hat{\alpha}$ when defining $\pm\sqrt{|\hat{\alpha}|}$. The
specific disambiguation depends on the physical meaning and use of the
resulting operator.

For instance, $\pm\sqrt{|\alpha|}$ could appear in a Dirac observable
$\phi\pm\sqrt{|\alpha|}$ of the constraint
$C=p_{\phi}^2-4|\alpha|p_{\alpha}^2$, defining relational evolution
$\phi(\alpha)$ by setting the observable to a constant value. (The constraint
surface has two components. For a given sign choice, in
$\phi\pm\sqrt{|\alpha|}$, a Dirac observable on one of the components is
obtained.) Not only quantization but also well-defined classical evolution
then requires a disambiguation. As in our main example, a suitable choice can
be derived by comparing relational evolution with the behavior on gauge
orbits, which is easier in this model because here $\alpha$ does not have a
turning point.

The gauge equations can easily be solved for
$\phi(\epsilon)=\phi_0+2p_{\phi}\epsilon$,
$|\alpha(\epsilon)|=4p_{\phi}\epsilon^2$ and
$p_{\alpha}(\epsilon)=-{\rm sgn}(\alpha)/(4\epsilon)$ while $p_{\phi}$ is
constant. Instead of a turning point, $\alpha(\epsilon)$ has a stationary
point at $\alpha=0$ where ${\rm d}\alpha/{\rm d}\epsilon=0$. Positive and
negative $\alpha$ therefore need not be connected to a single gauge
orbit. However, $\phi(\epsilon)$ clearly moves through all positive and
negative values. The relational description based on
$\phi\pm\sqrt{|\alpha|}={\rm constant}$ can correctly describe this behavior
only if the disambiguation
$\pm\sqrt{|\alpha|}={\rm sgn}(\alpha)\sqrt{|\alpha|}$ is used (up to an
overall sign choice).

The suitability of this disambiguation can also be seen
from the fact that the model system can be obtained by a canonical
transformation from the pair $(x,p_x)$ with constraint $p_{\phi}^2-p_x^2$ and
Dirac observable $\phi-x$ (on one component of the constraint surface). A
simple interpretation of $\pm\sqrt{|\alpha|}=\sqrt{|\alpha|}$, ignoring the
sign choice, would follow from a local canonical transformation
$x=\sqrt{|\alpha|}$, $p_x=2\sqrt{|\alpha|}p_{\alpha}$ that is not defined on
negative $x$. The global transformation $x={\rm sgn}{\alpha}\sqrt{|\alpha|}$,
$p_x=2\sqrt{|\alpha|}p_{\alpha}$ (up to an overall sign choice in $x$)
corresponds to our disambiguation.

Back to our main example, the unwinding of the local clock $\alpha$ to a
global time parameter $\tau$ is an example of a disambiguation, determined as
in the simple model by the condition that there should be states in which
$\phi$ has the expected asymptotic semiclassical behavior according to which
it goes to negative infinity at early times and positive infinity at late
times.  This variable classically has a strictly monotonic gauge flow for
non-zero $p_{\phi}$ but inherits double-valuedness when it is constructed from
the Dirac observable $\Phi$. The double-valuedness is resolved by the
disambiguation used throughout the paper.

In some cases, double-valuedness can be resolved in quantum mechanics if one
requires that states are always superpositions of states suitably supported on
both values. For instance, working only with even wave functions
$\psi(\alpha)$ treats positive and negative $\alpha$ on the same footing. For
operators to respect this symmetry, a specific sign choice must be made, such
as using the same expression $\sqrt{|\alpha|}$ for positive and negative
$\alpha$. In our case and in other models of relational evolution, however,
this procedure is not suitable for various reasons: (i) The symmetry of states
that helps to disambiguate possible operators is imposed by hand and not
derived from the dynamics. (ii) The condition is not guaranteed to be
consistent with the required semiclassical behavior in asymptotic
regimes. (iii) The model would be strongly restricted because any choice of
initial states would have to follow the imposed symmetry condition. For
instance, it would be impossible to set up a semiclassical state that compares
quantum evolution with a single classical trajectory going through a given
pair $(\alpha,\phi)$. (iv) As we have seen in our main discussion, the
symmetry condition that connects different $\alpha$ before and after a turning
point is usually not universal but, just like $\alpha_{\rm t}$, depends on
other phase-space degrees of freedom such as $p_{\phi}$. (v) The strict
symmetry condition ignores the possibility that quantum physics on phase
spaces with non-trivial topology, as implied here by branch cuts on the
solution space of the constraint, usually makes use of the universal covering
space on which discrete classical symmetries are not necessarily
respected. Accordingly, our solutions are not reflection symmetric around the
turning point, which may not be obvious from the plot but can easily be seen
from the fact that the Gaussian form of our initial state, posed on one side
of the turning point, is, generically, not recovered on the other side by the
non-harmonic evolution of our model.

\subsection{Dirac observables in evolution with respect to a local clock}

The standard procedure of relational evolution breaks down if one is
interested in developing a picture of quantum evolution in terms of $\alpha$
rather than $\phi$. Since $\hat{p}_{\phi}^2-e^{\gamma\hat{\alpha}}$ is not
positive definite, the definition of its square root is not clear. Even if
this problem can be solved, an exact factorization such as (\ref{Cpm}) is no
longer available because $\hat{p}_{\alpha}$ does not commute with
$\hat{p}_{\phi}^2-e^{\gamma\hat{\alpha}}$. It is possible to define the
constraint operator as
\begin{eqnarray}
  \hat{\mathcal{C}}&=&\left(\hat{p}_{\alpha}-\sqrt{\hat{p}_{\phi}^2-e^{\gamma\hat{\alpha}}}\right) 
                       \left(\hat{p}_{\alpha}+\sqrt{\hat{p}_{\phi}^2-e^{\gamma\hat{\alpha}}}\right)\nonumber\\
  &=&
\hat{p}_{\alpha}^2 - \hat{p}_{\phi}^2+ e^{\gamma\hat{\alpha}}
+\left[\hat{p}_{\alpha}, \sqrt{\hat{p}_{\phi}^2-e^{\gamma\hat{\alpha}}}\right] \label{Cpm2}
\end{eqnarray}
in this specific ordering, with a commutator term that indicates quantum corrections to the classical
expression. However, since the two factors do not commute, acting on a state
by $\hat{\mathcal{C}}\psi$, as in standard Dirac quantization, can give only
solutions with the negative sign of $p_{\alpha}$.

Our procedure makes use of the same factorization, but since we explicitly
include phases of time reversal when the clock $\alpha$ runs in the opposite
direction of time $\tau$, we are acting not only to the right (with a positive
$\alpha$-Hamiltonian $H=-p_{\alpha}$ from the second factor of (\ref{Cpm2})),
but also on the left when ${\rm d}\alpha/{\rm d}\tau<0$ where the
$\alpha$-Hamiltonian $H=-{\rm sgn}({\rm d}\alpha/{\rm d}\tau) p_{\alpha}>0$ is
still positive if we use the first factor of (\ref{Cpm2}). The
$p_{\phi}$-dependence of time reversals implied by the $p_{\phi}$-dependence
of $\alpha_{\rm t}$ means that the standard definition of a physical Hilbert
space does not apply. Nevertheless, we were able to define unitary
$\alpha$-evolution on a physical Hilbert space (defined as a Hilbert space on
which gauge degrees of freedom are not represented) with a conserved inner
product.

Moreover, our expectation value (\ref{expect}) faithfully models the Dirac
observable $\Phi$ far from the turning points. The $\tanh^{-1}$-part of
(\ref{expect}), evaluated in a semiclassical state, is equivalent to constant
$\Phi$ with a specific branch choice of the inverse cosh determined by
${\rm sgn}({\rm d}\alpha/{\rm d}\tau)$. The last term in (\ref{expect}) shows
an additional quantum correction starting at the turning point, thanks to the
theta function. This correction leads to stronger deviations between classical
and quantum behavior, such as the non-monotonic
$\langle\hat{\phi}\rangle(\tau)$ seen in Fig.~\ref{fig:phitaup} for larger
values of $p_{\phi}$.

Asymptotically, the term implies the quantum contribution $\Delta\phi_1$ to
the shift of $\phi$ derived in (\ref{Deltaphi}). Quantum evolution through the
turning point therefore connects two semiclassical evolutions with respect to
two different Dirac observables, $\Phi_1$ and
$\Phi_2=\Phi_1+\Delta\phi_1$. The classical Dirac observable $\Phi$ is not
strictly conserved, but since it is not globally defined it does not have a
direct quantum analog anyway. Our construction implies a successful
implementation of Dirac observables that are conserved at least in
semiclassical regimes far from turning points. As our analytical expressions
show, the overall shift $\Delta\phi =4\gamma^{-1}(\ln 2-1)$ between the
asymptotic past and the asymptotic future of a quantum trajectory decreases
for larger $\gamma$, in which case the exponential potential is steeper and
most $p_{\phi}$-contributions to a state reach their turning point at almost
the same time. Additional quantum corrections implied by the
$p_{\phi}$-dependence of the turning point are then suppressed.

\section{Conclusions}

We applied the methods of \cite{Gribov,LocalTime,Period} to a
simple cosmological model that, after rescalings and simplifications, has a
Hamiltonian constraint with a standard kinetic energy for two variables, and
an exponential potential for one of them. One of the degrees of freedom
therefore can be used as a global internal time, while the other one
encounters one turning point in its classical evolution and constitutes a
local clock variable. Traditional methods of deparameterization are able to
implement only the global time, while we obtained consistent quantum evolution
also with the local clock.

Unlike in previous proposals our quantum evolution does not freeze when the
classical turning point is reached, as we demonstrated both analytically by
equation (\ref{expect}) and numerically in Figs.~\ref{fig:phitau} and
\ref{fig:density1}. Our construction also gives a successful description of
the fact that a quantum superposition of different energy eigenstates
encounters a number of turning points at different times, in contrast to the
classical system at a fixed energy. A visible implication of this new behavior
can be seen in the possibility that the variable that could be used as a
global internal time may behave non-monotonically when it evolves with respect
to the local clock, as shown in Fig.~\ref{fig:phitaup}. The same feature
demonstrates that the two choices, global clock or local clock, are not
equivalent upon quantization, adding another example to the list of models
that have confirmed the inequivalence between different choices of internal
times in quantum cosmology
\cite{MultChoice,ClocksDyn,TwoTimes,SingClock,BianchiInternal,ClockDep}.
% >>>
(Our results are, however, invariant with respect to non-linear
reparameterizations of the effective time parameter $\tau$ if they are
performed after the local clock $\alpha$ has been chosen. The resolution of
the turning point and the characteristic shift therefore do not depend on the
parameterization (\ref{parameter}) as long as it presents a full
disambiguation of the flow of $\alpha$.)
% <<<

The energy dependence of classical turning points also implies a
characteristic shift between semiclassical relational trajectories in the
asymptotic past and future. The corresponding analytical expressions allowed
us to demonstrate that the classical Dirac observables are approximately
conserved asymptotically far from the turning point, but undergo specific
changes during the transition through a turning point. This feature is
remarkable because the corresponding Dirac observable does not have a
well-defined quantization in this case, owing to multi-valuedness of the
classical expression on phase space. (See also \cite{DiracChaos,DiracChaos2}
for a discussion of Dirac observables on non-trivial phase spaces.)
Multi-valuedness is related precisely to the behavior around turning points
that implies the non-conservation of classical Dirac observables at the
quantum level. Our procedure therefore preserves the expected classical
features in regimes where turning points are not relevant, and at the same
time provides a consistent freeze-free quantum evolution through turning
points.

While the dependence of the turning point on the energy of a state is a
characteristic feature, for a given energy the turning point is unique, given
the monotonic nature of the potential used here.  In this respect, the model
is rather different from the first application of our methods in
\cite{LocalTime}, where the clock variable was oscillating and encountered its
turning points an infinite number of times even in a single energy
eigenstate. The transition through the turning point remains hard to analyze
without numerical input because it is sensitive to the various energy
contributions of a wave packet that transit the turning point at different
times.  But our detailed investigation of a model with a single turning point
has led to a clear analytical description of the relationship between states
before and after the turning point, which might also help to understand
long-term quantum evolution in the presence of local oscillating clocks.

A common feature can be seen in the observation that quantum evolution with
local clocks is closer to what is expected from global deparameterization when
the potential is very steep in the range of variables where turning points
occur. For an oscillating clock $\phi$, this happened for large clock
frequencies in a potential $\lambda^2\phi^2$ with large $\lambda$. In the
present case, the condition is large $\gamma$, such that the potential
$e^{\gamma\alpha}$ for the local clock $\alpha$ with a single turning point
gets very steep.

\section{Acknowledgements}

We thank Philipp Hoehn, Rachael Huxford, Valerie Martinez, and Radu Roiban for their input
and contribution to discussions. This work was supported in part by the Sloan
Foundation and Penn State's Office for Educational Equity and by NSF grant PHY-2206591.

%\bibliographystyle{../preprint}
%\bibliography{../Bib/QuantGra.bib}

\begin{thebibliography}{10}

\bibitem{GenHamDyn1}
P.~A.~M.\ Dirac,
\newblock Generalized Hamiltonian dynamics,
\newblock {\em Can.\ J.\ Math.} 2 (1950) 129--148

\bibitem{QuantumRef1}
F.\ Giacomini, A.\ Castro-Ruiz, and C.\ Brukner,
\newblock Quantum mechanics and the covariance of physical laws in quantum
  reference frames,
\newblock {\em Nat.\ Commun.} 10 (2019) 494, [arXiv:1712.07207]

\bibitem{QuantumRef2}
A.\ Vanrietvelde, P.~A.\ Hoehn, F.\ Giacomini, and E.\ Castro-Ruiz,
\newblock A change of perspective: switching quantum reference frames via a
  perspective-neutral framework,
\newblock {\em Quantum} 4 (2020) 225, [arXiv:1809.00556]

\bibitem{QuantumRef3}
A.\ Vanrietvelde, P.~A.\ Hoehn, and F.\ Giacomini,
\newblock Switching quantum reference frames in the $N$-body problem and the
  absence of global relational perspectives, [arXiv:1809.05093]

\bibitem{QuantumRef4}
P.~A.\ Hoehn, A.~R.~H.\ Smith, and M.~P.~E.\ Lock,
\newblock The Trinity of Relational Quantum Dynamics,
\newblock {\em Phys.\ Rev.\ D} 104 (2021) 066001, [arXiv:1912.00033]

\bibitem{QuantumRef5}
P.~A.\ Hoehn, A.~R.~H.\ Smith, and M.~P.~E.\ Lock,
\newblock Equivalence of approaches to relational quantum dynamics in
  relativistic settings,
\newblock {\em Front.\ Phys.} 9 (2021) 587083, [arXiv:2007.00580]

\bibitem{QuantumRef6}
F.\ Giacomini,
\newblock Spacetime Quantum Reference Frames and superpositions of proper
  times,
\newblock {\em Quantum} 5 (2021) 508, [arXiv:2101.11628]

\bibitem{Blyth}
W.~F.\ Blyth and C.~J.\ Isham,
\newblock Quantization of a Friedmann universe filled with a scalar field,
\newblock {\em Phys.\ Rev.\ D} 11 (1975) 768--778

\bibitem{BrownKuchar}
J.~D.\ Brown and K.~V.\ Kucha\v{r},
\newblock Dust as a standard of space and time in canonical quantum gravity,
\newblock {\em Phys.\ Rev.\ D} 51 (1995) 5600--5629

\bibitem{HusainDust}
V.\ Husain and T.\ Pawlowski,
\newblock Dust reference frame in quantum cosmology,
\newblock {\em Class.\ Quantum Grav.} 28 (2011) 225014, [arXiv:1108.1147]

\bibitem{WaldTimeModels}
A.\ Higuchi and R.~M.\ Wald,
\newblock Applications of a new proposal for solving the `problem of time' to
  some simple quantum cosmological models,
\newblock {\em Phys.\ Rev.\ D} 51 (1995) 544--561, [gr-qc/9407038]

\bibitem{PhysEvolBI}
M.\ Mart{\'\i}n-Benito, G.~A.\ Mena~Marug\'an, and T.\ Pawlowski,
\newblock Physical evolution in Loop Quantum Cosmology: The example of vacuum
  Bianchi I,
\newblock {\em Phys.\ Rev.\ D} 80 (2009) 084038, [arXiv:0906.3751]

\bibitem{Gribov}
M.~M.\ Amaral and M.\ Bojowald,
\newblock A path-integral approach to the problem of time,
\newblock {\em Ann.\ Phys.} 388C (2018) 241--266, [arXiv:1601.07477]

\bibitem{LocalTime}
G.\ Wendel, L.\ Mart\'{\i}nez, and M.~Bojowald,
\newblock Physical implications of a fundamental period of time,
\newblock {\em Phys.\ Rev.\ Lett.} 124 (2020) 241301, [arXiv:2005.11572]

\bibitem{Period}
M.\ Bojowald, L.\ Mart\'{\i}nez, and G.\ Wendel,
\newblock Relational evolution with oscillating clocks,
\newblock {\em Phys.\ Rev.\ D} 105 (2022) 106020, [arXiv:2110.07702]

\bibitem{Infrared}
M.\ Bojowald,
\newblock The BKL scenario, infrared renormalization, and quantum cosmology,
\newblock {\em JCAP} 01 (2019) 026, [arXiv:1810.00238]

\bibitem{GenRepIn}
J.~B.\ Hartle and D.\ Marolf,
\newblock Comparing Formulations of Generalized Quantum Mechanics for
  Reparametrization-Invariant Systems,
\newblock {\em Phys.\ Rev.\ D} 56 (1997) 6247--6257, [gr-qc/9703021]

\bibitem{MultChoice}
P.\ Malkiewicz,
\newblock Multiple choices of time in quantum cosmology,
\newblock {\em Class.\ Quantum Grav.} 32 (2015) 135004, [arXiv:1407.3457]

\bibitem{ClocksDyn}
P.\ Malkiewicz,
\newblock Clocks and dynamics in quantum models of gravity,
\newblock {\em Class.\ Quantum Grav.} 34 (2017) 145012, [arXiv:1601.04857]

\bibitem{TwoTimes}
M.\ Bojowald and T.\ Halnon,
\newblock Time in quantum cosmology,
\newblock {\em Phys.\ Rev.\ D} 98 (2018) 066001, [arXiv:1612.00353]

\bibitem{SingClock}
S.\ Gielen and L.\ Men\'{e}ndez-Pidal,
\newblock Singularity resolution depends on the clock,
\newblock {\em Class.\ Quantum Grav.} 37 (2020) 205018, [arXiv:2005.05357]

\bibitem{BianchiInternal}
P.\ Malkiewicz, P.\ Peter, and S.~D.~P.\ Vitenti,
\newblock Quantum empty Bianchi I spacetime with internal time,
\newblock {\em Phys.\ Rev.\ D} 101 (2020) 046012, [arXiv:1911.09892]

\bibitem{ClockDep}
S.\ Gielen and L.\ Men\'{e}ndez-Pidal,
\newblock Unitarity, clock dependence and quantum recollapse in quantum
  cosmology,
\newblock {\em Class.\ Quantum Grav.} 39 (2022) 075011, [arXiv:2109.02660]

\bibitem{DiracChaos}
B.\ Dittrich, P.~A.\ Hoehn, T.~A.\ Koslowski, and M.~I.\ Nelson,
\newblock Chaos, Dirac observables and constraint quantization,
  [arXiv:1508.01947]

\bibitem{DiracChaos2}
B.\ Dittrich, P.~A.\ Hoehn, T.~A.\ Koslowski, and M.~I.\ Nelson,
\newblock Can chaos be observed in quantum gravity?,
\newblock {\em Phys.\ Lett.\ B} 769 (2017) 554--560, [arXiv:1602.03237]

\end{thebibliography}

\end{document}